\documentclass[reprint, superscriptaddress, amsmath, amssymb, aps, prx]{revtex4-2}

\usepackage{pifont}
\usepackage{graphicx}
\usepackage{bm}
\usepackage{sidecap}
\usepackage{xcolor}
\usepackage[colorlinks=true,citecolor=blue]{hyperref}
\usepackage{natbib}
\usepackage{amsmath}
\usepackage{amssymb}
\usepackage{comment}
\usepackage{bm}

\begin{document}

\title{Mirror-Symmetry-Enforced Photonic Altermagnet}
  
\author{Chong Cao}
\affiliation{School of Physics and Optoelectronics, Xiangtan University, Xiangtan 411105, China}

\author{Xiong-Xiong Xue}
\affiliation{School of Physics and Optoelectronics, Xiangtan University, Xiangtan 411105, China}

\author{Yee Sin Ang}
\email{yeesin\_ang@sutd.edu.sg}
\affiliation{Science, Mathematics and Technology (SMT), Singapore University of Technology and Design, Singapore 487372}

\author{Haiyu Meng}
\email{phymhy@xtu.edu.cn}
\affiliation{School of Physics and Optoelectronics, Xiangtan University, Xiangtan 411105, China}

\begin{abstract}

Altermagnets host momentum-dependent spin splitting without net magnetization, a symmetry-enforced band phenomenon whose photonic analogues have so far been realized only in square lattices governed by fourfold rotation. Here we introduce a photonic altermagnet on a hexagonal lattice whose helicity splitting is governed by mirror rather than rotational symmetry. Elliptical chiral elements of alternating handedness, placed at the vertices of a regular hexagon, leave the two opposite-chirality sublattices connected only by chirality reversal combined with a mirror reflection. Full-wave simulations reveal mirror-related splitting of the two opposite-helicity branches in the band structure and isofrequency contours, with the channels exchanged when the ellipse orientation is reversed. Using a finite photonic crystal slab, we show that such splitting separates a linearly polarized beam into handedness-resolved channels, thus enabling beam splitting and direction-selective helicity filtering with target-helicity output fractions above 0.85 and output paths continuously tunable through the ellipse rotation angle. These results extend photonic altermagnetism to a previously unexplored lattice-symmetry class and establish mirror-symmetric chiral textures as building blocks for altermagnetism-inspired on-chip chiral photonics.

\end{abstract}

\maketitle

\section{Introduction}

Altermagnetism is an emerging class of collinear magnetism that combines the spin-split bands of ferromagnets with the vanishing net magnetization of antiferromagnets~\cite{Smejkal2022a,Smejkal2022b,bai2024altermagnetism,song2025altermagnets,liu2026symmetry,zeng2026classification,zhang2026arpes,cheong2025altermagnetism,jungwirth2026symmetry}. Its defining feature is a momentum-dependent, even-parity ($d$-, $g$-, or $i$-wave) lifting of the spin degeneracy, enforced not by relativistic spin--orbit coupling but by the spin-group symmetry that relates the opposite-spin sublattices through a crystalline rotation rather than translation or inversion ~\cite{Smejkal2022a,Smejkal2020,jiang2025metallic,wang2024electric,chen2023giant,che2024realizing,zeng2024description,naka2025altermagnetic}. This symmetry-driven mechanism has made altermagnetism a focus of condensed-matter research, underlying phenomena such as the crystal and anomalous Hall effects~\cite{yu2025neel,zhou2025manipulation,liu2025anomalous,jin2024anomalous,takahashi2025elasto,sheoran2025spontaneous,sato2024altermagnetic}, the spin-splitter torque~\cite{liu2026altermagnetic,zhang2025electrical,guo2024direct,han2024electrical,cui2023efficient,vakili2025spin}, tunneling magnetoresistance~\cite{fang2026one,yang2026altermagnetic,samanta2025spin,samanta2024tunneling,jiang2023prediction,wang2026pentagonal,yang2025unconventional,li2025tunneling,noh2025tunneling,liu2024giant} and unconventional superconducting pairing~\cite{Gonzalez2021,Bai2022,fukaya2025josephson,zhang2024finite,ouassou2023dc}. Altermagnetism is further enriched by multiferroicity and other ferroic-order couplings~\cite{Gu2025,peng2026sliding,sun2026unified,duan2025antiferroelectric,peng2025ferroelastic,ding2025ferroelastically} and van der Waals and bilayer-stacking engineering~\cite{Liu2024twist,Pan2024,Hodt2024,zhu2025two,zhu2026altermagnetic,peng2025all,zeng2024bilayer}. In electronic crystals, however, it remains limited by a small pool of firmly established hosts and by the difficulty of confirming the order: the archetypal candidate RuO$_2$, for instance, is still debated, with muon-spin-rotation and neutron studies pointing to a nonmagnetic ground state even as transport experiments report altermagnetic-like signatures~\cite{RuO2review,RuO2torque}.

Photonic crystals provide a versatile platform that sidesteps these material constraints. By supporting Bloch bands and crystalline symmetry representations that closely mirror those of electronic solids, photonic crystals further offer internal degrees of freedom -- polarization, helicity, and orbital character -- that are more readily engineered than electronic spin~\cite{Joannopoulos2008,Lu2014,Bliokh2015}. Recent advances in planar dielectric structures also show that optical chirality can be strongly engineered in photonic platforms~\cite{gorkunov2025substrate}. Through the correspondence between electron spin and photon helicity, and between magnetic moment and material chirality, the spin-group symmetries underlying altermagnetism map onto photonic lattices, with chirality reversal playing the role of magnetic time reversal~\cite{Kim2025,Qiu2026}. Recent pilot studies have demonstrated $d$-wave helicity-split bands in a chiral photonic crystal~\cite{Kim2025} and $d_{xy}$-wave pseudospin splitting in an orbital altermagnetic photonic crystal ~\cite{Qiu2026}. Beyond reproducing the altermagnetic band structure, these \emph{altermagnetic photonic crystals} \textcolor{blue}{point} towards useful functionalities such as helicity-contrasting beam splitting and filtering. Crucially, the geometry, lattice symmetry, and chirality distribution of a photonic crystal can be engineered almost at will, and the scale invariance of Maxwell's equations enables the operation frequency to be conveniently rescaled from microwave to optical frequencies.

\begin{figure*}[t]
\centering
\includegraphics[width=\textwidth]{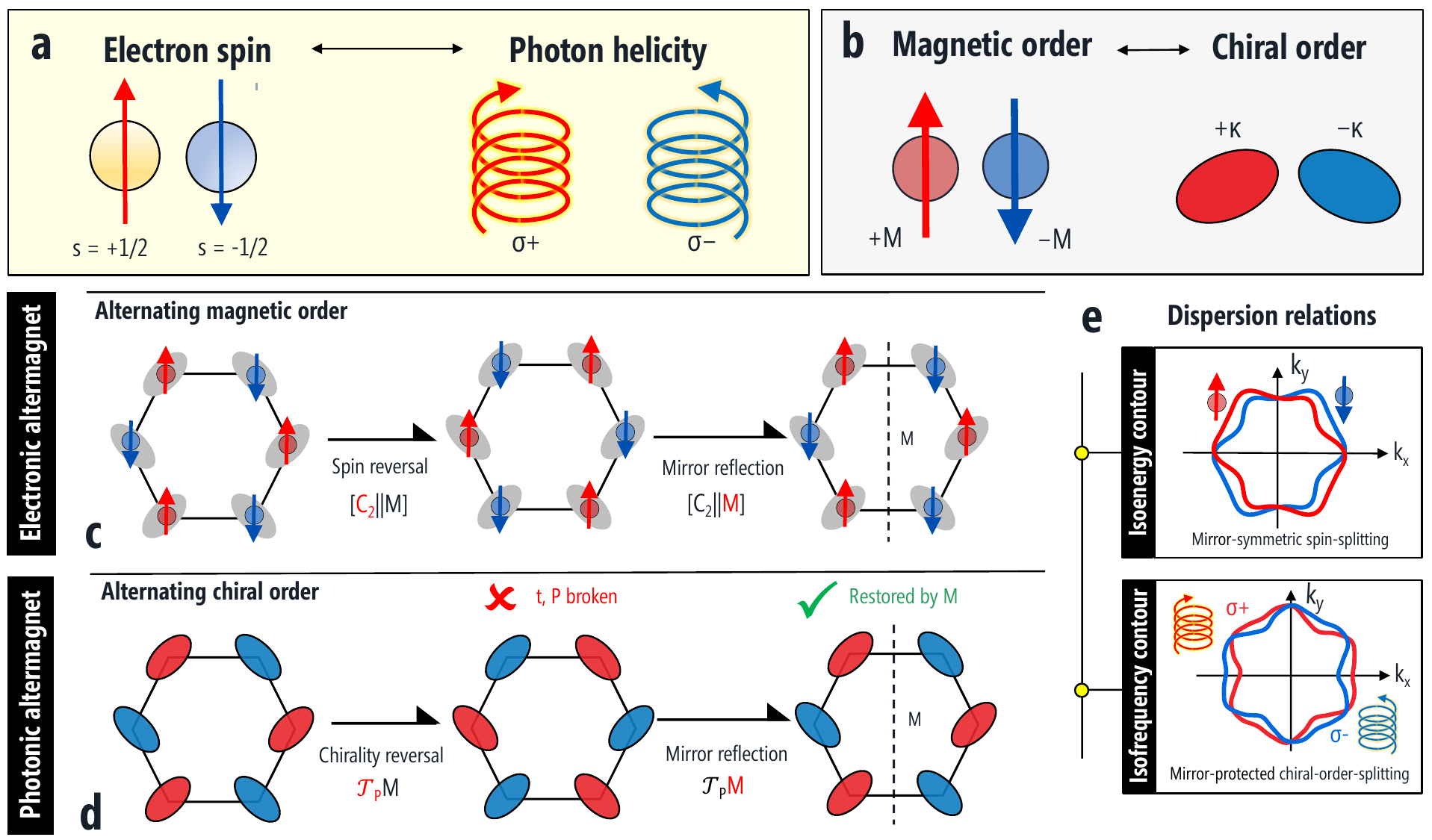}
\caption{Symmetry correspondence between electronic altermagnetism and a hexagonal chiral elliptical photonic crystal. (a) Correspondence between electronic spin and photon helicity. The two spin states ($s=+1/2$) and ($s=-1/2$) are mapped to the two photonic helicity channels ($\sigma_{+}$) and ($\sigma_{-}$), respectively, with red and blue indicating opposite spin or helicity states.
(b) Analogy between magnetic and chiral orders. Magnetic units carrying opposite out-of-plane magnetic moments, ($+M$) and ($-M$), are represented in the photonic system by optically active elements with opposite material chirality, ($+\kappa$) and ($-\kappa$).
(c) Symmetry operation in an electronic altermagnetic configuration. After spin reversal, the magnetic pattern is not restored by the reversal operation alone, but through a subsequent mirror operation ($M$), leading to spin splitting under zero net magnetization.
(d) Same as (c) but for photonic altermagnet. Here chirality reversal ($\mathcal{T}_{p}$) exchanges the +$\kappa$ and -$\kappa$ elements, while the mirror operation relates the reversed chiral texture to the original configuration. 
(e) Comparison of momentum-space spin splitting, showing the mirror-constrained spin-split isoenergy contours in an electronic altermagnet (upper panel) and  the helicity-split IFCs in the photonic altermagnet (lower panel).}
\label{fig}
\end{figure*}

While prior demonstrations of photonic altermagnets are confined to square lattices in which the sublattice-connecting operation is a fourfold rotation, i.e.\ the unitary $C_4$ or the antiunitary $C_{4z}\mathcal{T}$ ~\cite{Kim2025,Qiu2026}, whether altermagnetism can be realized beyond square symmetry remains an open question. In this work, we demonstrate altermagnetism in a chiral photonic crystal whose helicity splitting is driven by mirror rather than rotational symmetry. With six alternating-handedness elliptical elements placed at the vertices of a regular hexagon and truncated by the cell boundary, the two opposite-chirality sublattices are connected not by chirality reversal, a $\mathcal{PT}$-type operation, or translation, but only by chirality reversal combined with a mirror reflection---identifying the mirror as the symmetry operation that enforces the helicity splitting. Using full-wave simulations, we compute the helicity-resolved band structure and isofrequency curves (IFCs) and find mirror-related splitting of the two opposite-helicity (circular-polarization) branches. Notably, the momentum-space splitting enables handedness-dependent beam splitting and direction-selective helicity filtering, with target-helicity output fractions exceeding $0.85$, and the output direction and channel separation can be continuously tuned through the ellipse rotation angle. These findings extend the photonic realization of altermagnetism to a previously unexplored lattice-symmetry class and establish mirror-symmetric chiral textures as building blocks for altermagnetism-inspired chiral photonics compatible for compact on-chip integrations.

\section{Model and Symmetry Analysis}

A photonic altermagnet must split the two opposite-helicity channels in momentum space while keeping the net chiral response zero, just as an electronic altermagnet obtains spin splitting from compensated magnetic sublattices whose spatial arrangement breaks the effective spin-reversal symmetry at general momenta. We therefore construct a chiral photonic structure with balanced positive and negative chirality $\pm\kappa$, but with a spatial arrangement that removes the global degeneracy between the two helicity channels.

The corresponding relation between electronic-magnetic order and photonic chiral order is illustrated in Figs. 1(a) and 1(b). First,  photon helicity states carry spin angular momentum and so play the role of electron spin. Second, optical elements with opposite material chirality, $+\kappa$ and $-\kappa$, mimic magnetic units with opposite moments, ($+M$) and ($-M$). Third, chirality reversal in the photonic system corresponds to magnetic-moment reversal in the electronic system. In this sense, a uniform chiral arrangement is analogous to a ferromagnetic order, whereas an alternating chiral arrangement resembles an antiferromagnetic order. However, a photonic altermagnet requires more than alternating chirality: the two opposite-chirality substructures must not be restorable by chirality reversal or translation alone. The coupling between material chirality and photon helicity follows from the constitutive relations of an isotropic Pasteur medium. With the time convention $e^{-i\omega t}$~\cite{Serdyukov2001}, they are written as
\begin{equation}
\mathbf{D}=\varepsilon \mathbf{E}+i\frac{\kappa}{c_{0}}\mathbf{H} \qquad
\mathbf{B}=\mu \mathbf{H}-i\frac{\kappa}{c_{0}}\mathbf{E}
\end{equation}
Here, $\kappa$ is the chirality parameter and ($c_0$) is the speed of light in vacuum. Together with the source-free Maxwell equations, 
\begin{equation}
\nabla \times \mathbf{E}=i\omega \mathbf{B} \qquad
\nabla \times \mathbf{H}=-i\omega \mathbf{D}
\end{equation}
the $\kappa$-dependent terms couple the electric and magnetic fields and make the eigenmodes sensitive to the handedness of the medium. Reversing the sign of $\kappa$ exchanges the responses of the two opposite helicity channels. Therefore, the spatial distribution of $pm\kappa$ determines whether the two helicity branches remain degenerate or become split in momentum space.

Guided by this correspondence, we construct a two-dimensional (2D) regular hexagonal cell containing six chiral elliptical elements located at the vertices, as shown in Fig.~1(d). The side length of the hexagon is $a_{0}=1~\mu\mathrm{m}$ and the axial ratio of each ellipse is $\alpha=1.3$. The background medium is set to $\varepsilon_{0}=\mu_{0}=1$, whereas the elliptical elements have $\varepsilon_{1}=\mu_{1}=2$. The red and blue ellipses carry opposite chirality parameters, $\kappa_{\mathrm{red}}=+1.5$ and $\kappa_{\mathrm{blue}}=-1.5$, respectively, and are arranged alternately around the hexagon. Because the numbers of $\kappa$ and $-\kappa$ elements are equal, the average chirality of the unit cell vanishes. The common $\varepsilon=\mu$ ratio further preserves the electromagnetic duality condition, allowing the helicity components to be well defined.

The key design feature is that the sign of $\kappa$ and the major-axis orientation of each ellipse jointly form a directional chiral texture. If only the material chirality is considered, the red and blue regions are fully compensated. They consequently cannot be superimposed by chirality reversal, by a $\mathcal{PT}$-type operation, or by translation alone, which lifts the global helicity degeneracy at general momenta. They are restored only by chirality reversal combined with a mirror reflection [Figs.~1(c)--1(d)], in direct analogy to the spin-reversal-plus-mirror operation that relates the sublattices of the electronic altermagnet in Fig.~1(c). This residual mirror symmetry constrains the splitting: because translation and $\mathcal{PT}$ symmetry are broken, the helicity channels are free to split in momentum space, yet because the mirror is preserved, the splitting must appear along mirror-related directions, as illustrated in Fig.~1(e). In this sense the crystal is globally achiral: although each element is locally chiral, the compensated $\pm\kappa$ populations and the mirror symmetry leave no net handedness, so the helicity splitting is altermagnetic in origin rather than a net-chirality effect.

To identify the helicity character of each eigenmode, we use the Riemann-Silberstein (RS) representation~\cite{BialynickiBirula}, which combines the electric field with the impedance-normalized magnetic field into two opposite-helicity components. In a homogeneous chiral medium the two helicities acquire distinct effective indices,
\begin{equation}
n_{\sigma}=\sqrt{\varepsilon\mu}+\sigma\kappa \qquad (\sigma=\pm1),
\end{equation}
illustrating the selective action of $\kappa$. For the periodic texture studied here, the bands, IFCs, and group velocities follow from full Bloch eigenvalue calculations. We solve $E_{z}$ and $H_{z}$ simultaneously in the 2D unit cell under Bloch boundary conditions, classify each mode by the helicity-polarization criterion defined below, and color the bands and contours accordingly. In this work, red and blue carry two distinct meanings: (i) on the structure diagrams, they mark the sign of the material chirality; (ii) on the band structures, IFCs, and propagation profiles, they mark the dominant eigenmode helicity. Finite-structure propagation is then computed under Gaussian-beam incidence, with the output helicity quantified by the normalized helicity-intensity difference.

\section{Results and Discussions}

\subsection{Helicity Splitting Induced by Mirror Symmetric Chiral Texture}

We first examine how the mirror-symmetric chiral texture lifts the degeneracy between the two helicity channels. To distinguish the helicity splitting from trivial effects caused by changes in the average filling fraction or average refractive index, we fix the positions of the six elliptical elements, $\varepsilon$, $\mu$, and $\kappa$, and change only the orientations of the ellipse major axes. Taking the red ellipse as the reference, clockwise rotation is defined as positive and counterclockwise rotation is defined as negative.

\begin{figure*}[t]
\centering
\includegraphics[width=\textwidth]{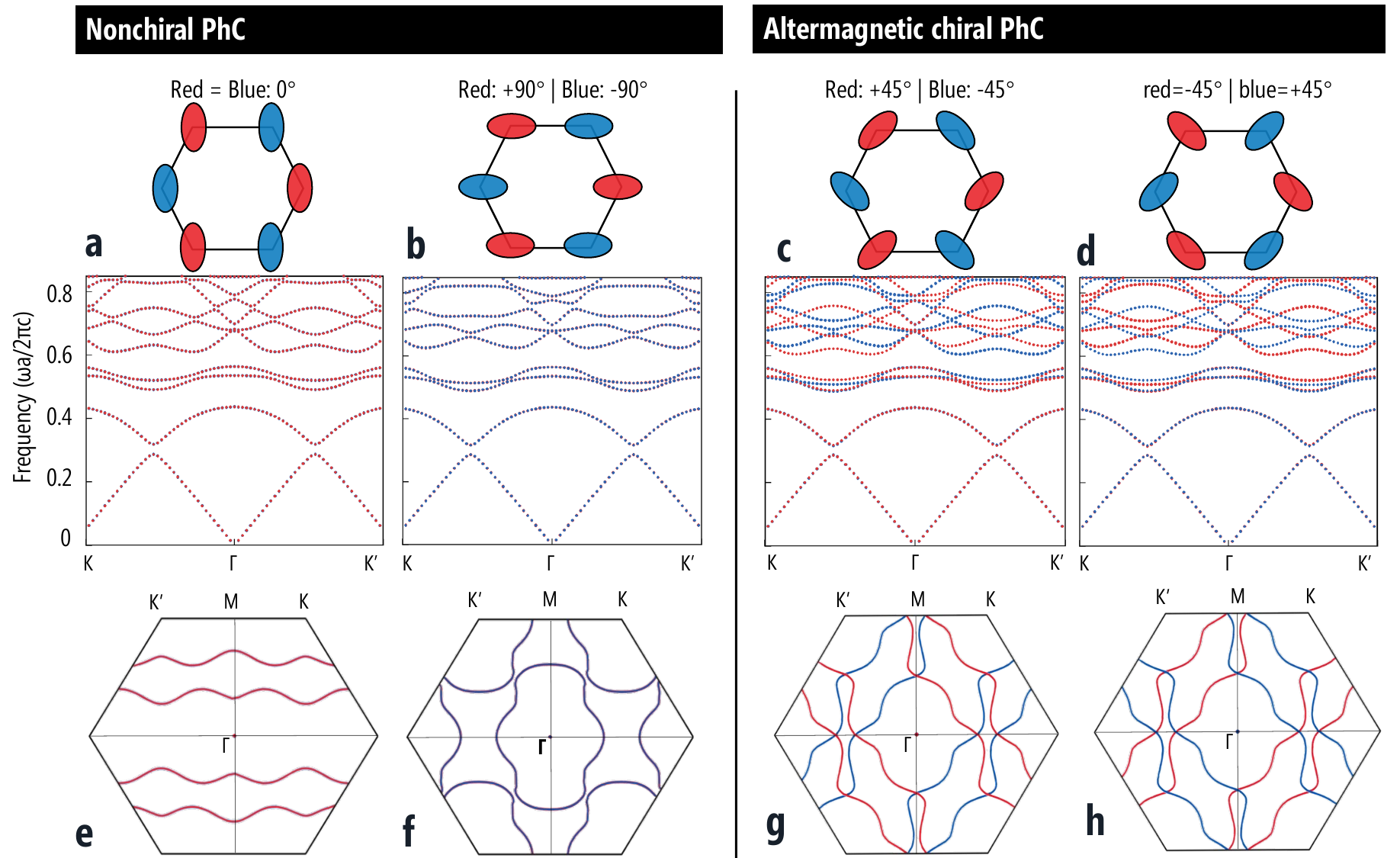}
\caption{
Control of the helicity response of the altermagnetic hexagonal photonic crystal by ellipse orientation.
Structures and corresponding band structures for four different ellipse orientations: (a) both the red and blue ellipses are oriented at $0^{\circ}$; (b) the red and blue ellipses are rotated by $+90^{\circ}$ and $-90^{\circ}$, respectively; (c) the red and blue ellipses are rotated by $+45^{\circ}$ and $-45^{\circ}$, respectively; and (d) the red and blue ellipses are rotated by $-45^{\circ}$ and $+45^{\circ}$, respectively.
The red and blue bands in the band diagrams represent two opposite helicity branches.
Panels (e) to (h) show the IFCs in reciprocal space extracted at the normalized frequency of 0.676 for the four structures shown in panels (a) to (d), respectively.
The red and blue curves represent the IFCs of the two opposite helicity branches.
The $+45^{\circ}$ and $-45^{\circ}$ ellipse orientation in (g), and the $-45^{\circ}$ and $+45^{\circ}$ ellipse orientation in (h), show clear interlaced patterns in both the band structures and the IFCs, whereas the $0^{\circ}$ and $\pm90^{\circ}$ ellipse orientations are helicity degenerate.
}
\label{fig:P2}
\end{figure*}

\begin{figure*}[t]
\centering
\includegraphics[width=\textwidth]{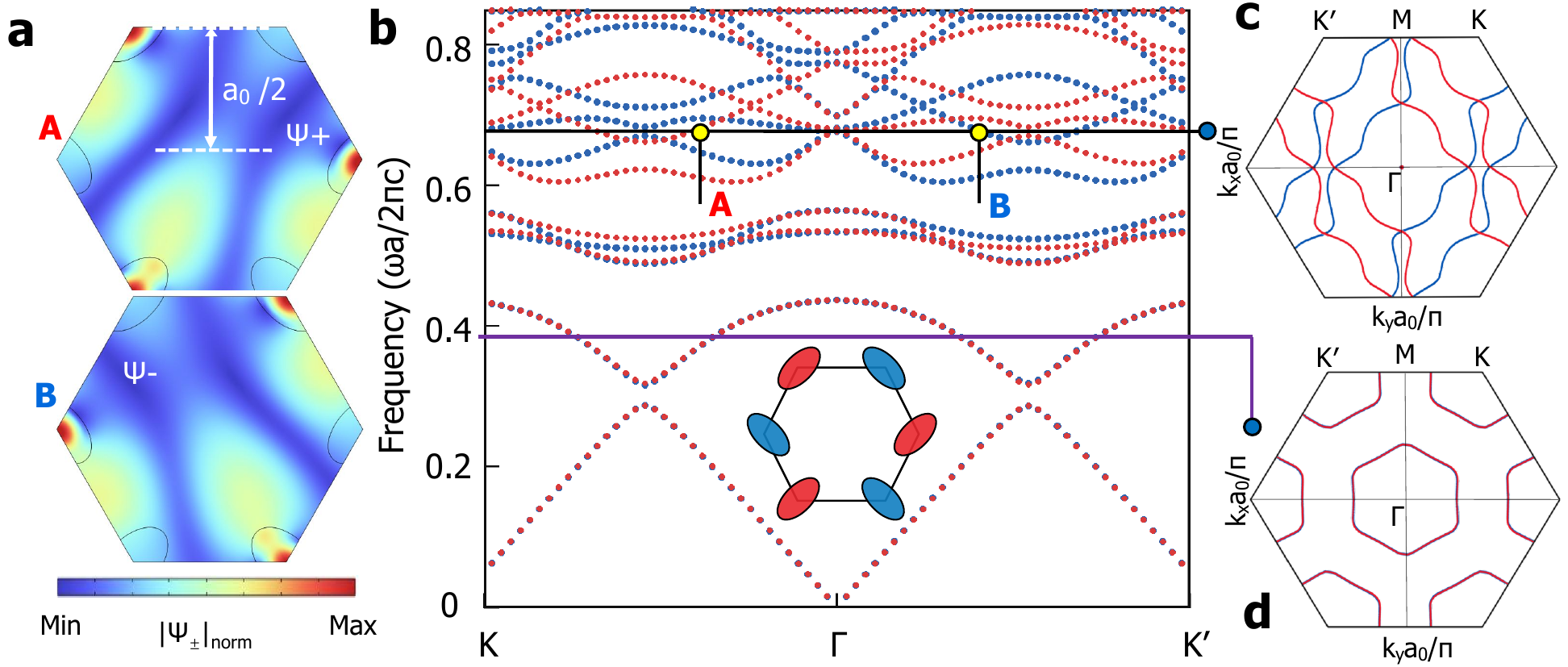}
\caption{
Helicity-polarized band structure and IFC in altermagnetic hexagonal photonic crystal.
(a) Normalized Riemann-Silberstein field distributions of two eigenmodes with opposite helicities in the hexagonal structure.
The upper and lower field maps correspond to the $\psi_{+}$ and $\psi_{-}$ components, respectively, and the colormap represents the normalized field intensity varying from the minimum value to the maximum value.
The marked A and B indicate the positions of the selected eigenmodes in the helicity-resolved band structures in (b).
The inset shows the arrangement of the six chiral elliptical units in the hexagonal unit cell. IFCs extracted at two representative normalized frequencies. of 0.676 (c), and of 0.38 (d).
}
\label{fig:P3}
\end{figure*}

Figure 2 compares four representative configurations. To determine the optical helicity of the bands, we analyze the eigenfields using the RS components \(\psi_{\pm}\), which correspond to opposite photonic helicities ~\cite{Kim2025},
\begin{equation}
\psi_{\pm}=\frac{E_z \pm i\eta H_z}{\sqrt{2}}
\label{eq:psi_plus}
\end{equation}
where $E_{z}$ and $H_{z}$ denote the $z$ component of the electric and magnetic fields, respectively, and $\eta H_{z}$ is the impedance-normalized magnetic field, which places the electric and magnetic components on the same scale. The degree of helicity polarization (DOHP) of each Bloch eigenmode can then be calculated by field integration over the basic hexagonal periodic region:
\begin{equation}
\mathrm{DOHP}=
\frac{\int_{\Omega} |\psi_{+}|^{2} \, \mathrm{d}A-\int_{\Omega} |\psi_{-}|^{2} \, \mathrm{d}A}
{\int_{\Omega} |\psi_{+}|^{2} \, \mathrm{d}A+\int_{\Omega} |\psi_{-}|^{2} \, \mathrm{d}A}
\end{equation}
Here, $\Omega$ denotes the basic hexagonal periodic region used for the integration of the eigenmode. A positive (negative) DOHP indicates dominance of the $\psi_{+}$ ($\psi_{-}$) component; $|\textrm{DOHP}|\to 1$ corresponds to a nearly single-helicity eigenmode, whereas $\textrm{DOHP}\to 0$ indicates strong mixing of the two helicities. The calculated DOHP is marked as red and blue in the band structures and IFC in Fig. 2.

When both the red and blue ellipses are oriented along $0^{\circ}$, the band structure along $K$--$\Gamma$--$K^{\prime}$ shows that the two helicity-dependent bands remain nearly degenerate [Fig. 2(a)]. Consistently, the corresponding IFC at $f=0.676$ does not exhibit a clear separation between the two helicity branches [Fig. 2(e)]. A similar behavior is found for the $90^{\circ}$ configuration [Fig. 2(b) and 2(f)], where the material parameters remain identical but the red and blue contours are still not sufficiently displaced to support strong helicity-dependent propagation. These two cases thus indicate that an alternating chiral distribution alone is insufficient to produce pronounced helicity splitting. 

In contrast, when the red and blue ellipses are oriented at $\pm45^{\circ}$ [Fig. 2(c)], respectively, the band structures split into two distinctive branches of opposite helicity, yielding a helicity-contrasting IFC [Fig. 2(g)]. Reversing the orientations between $\mp45^{\circ}$ exchanges the red and blue helicity branches [Fig. 2(d) and 2(h)]. Since all material parameters and the spatial positions of the ellipses are unchanged in Figs. 2(g) and 2(h), the exchange of the helicity branches cannot be attributed to a material-chirality effect or to an average-index perturbation, but reflects, instead, the directional coupling between the sign of $\kappa$ and the orientation-dependent scattering response as introduced by the elliptical major axis.

The stronger helicity splitting in the $45^{\circ}$ configurations arises from the interplay between the elliptical axes and the mirror-symmetric hexagonal lattice. For the $0^{\circ}$ and $90^{\circ}$ configurations, the major axes are nearly aligned with the principal structural directions, so the directional contrast experienced by the Bloch modes is weak. In contrast, the oblique $45^{\circ}$ orientations make the positive-and negative-chirality elements act as anisotropic chiral scatterers with distinct directional responses. The Bloch modes therefore resolve not only the sign of material chirality, but also the orientation-dependent scattering of each chiral element. This geometry-chirality coupling produces a $\mathbf{k}$-dependent helicity splitting while maintaining the mirror-related correspondence between the two branches.

In Fig. 3(a), we analyze representative eigenmodes, marked as $A$ and $B$ in the band structures in Fig. 3(b), to illustrate the spatial distributions of the normalized RS components when $\psi_{\pm}$ is dominant. Here, states $A$ and $B$ are dominated by $\psi_{+}$ and $\psi_{-}$, respectively. Since $\psi_{\pm}$ contains both the amplitude and phase information of $E_{z}$ and $\eta H_{z}$, the helicity polarization cannot be identified from $|E_{z}|$ alone. Instead, the electric and magnetic fields must be solved simultaneously, and the intensities of the two RS field components must be compared under the same normalization condition. Importantly, the band structure in Fig.~3(b) is helicity-split or helicity-degenerate depending on frequency: the bands are strongly split near $f=0.676$ [Fig. 3(c)] but remain nearly degenerate near $f=0.38$ [Fig.~3(d)]. As shown in the following sections, such band splitting enables linearly polarized beam to be spatially separated into two separate helicity channels ($\psi_\pm$). Such \emph{helicity-dependent} transport is frequency dependent and follows directly from the contour of the IFC.

\begin{figure*}[t]
\centering
\includegraphics[width=\textwidth]{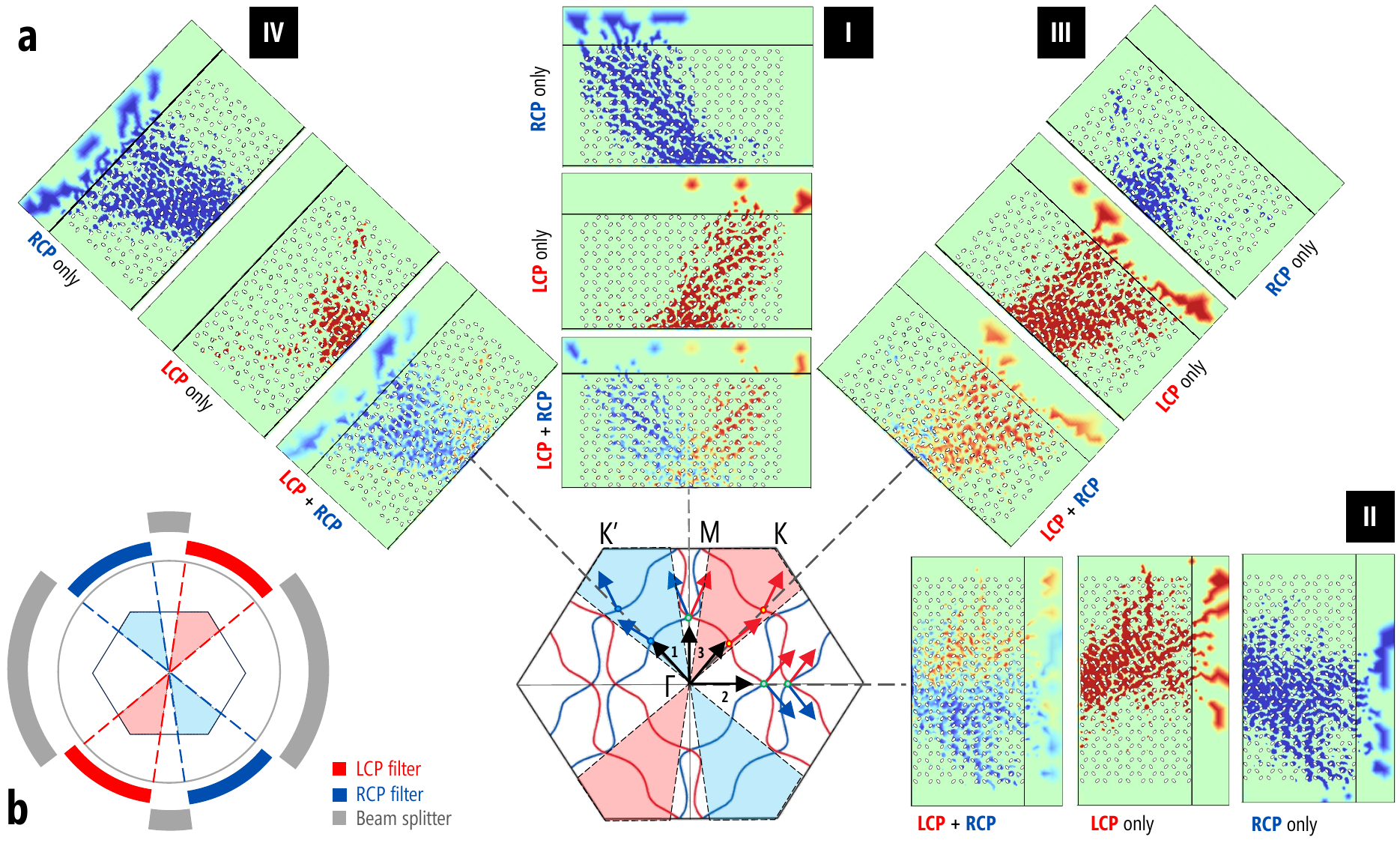}
\caption{
Helicity dependent beam splitting and filtering in the altermagnetic hexagonal photonic crystal.
(a) IFCs at the normalized frequency $f=0.676$ and beam propagation profiles under different incident directions relative to the hexagonal lattice.
The red and blue curves in the IFCs correspond to the two opposite helicity branches.
The black arrows denote the incident wave vector directions, and the colored arrows mark the group velocity directions associated with different helicity branches.
When the incident direction lies in the beam splitting regions labeled I and II, the two helicity components are simultaneously coupled into the photonic crystal and propagate along different directions, forming spatially separated red and blue outputs.
The corresponding beam profiles under single circular polarization incidence, labeled LCP only and RCP only, agree with the red and blue splitting channels obtained under linearly polarized incidence.
For the filtering directions labeled III and IV, only one helicity component can be effectively transmitted.
(b) Schematic representation of the beam manipulation functionality of the photonic crystal for different incident direction ranges.
The gray region denotes the beam splitting region, where the linearly polarized incident beam can be split into two opposite helicity components.
The red and blue regions denote the filtering regions, where the LCP and RCP components are mainly transmitted, respectively.
}
\label{fig:P4}
\end{figure*}

\begin{figure*}[t]
\centering
\includegraphics[width=\textwidth]{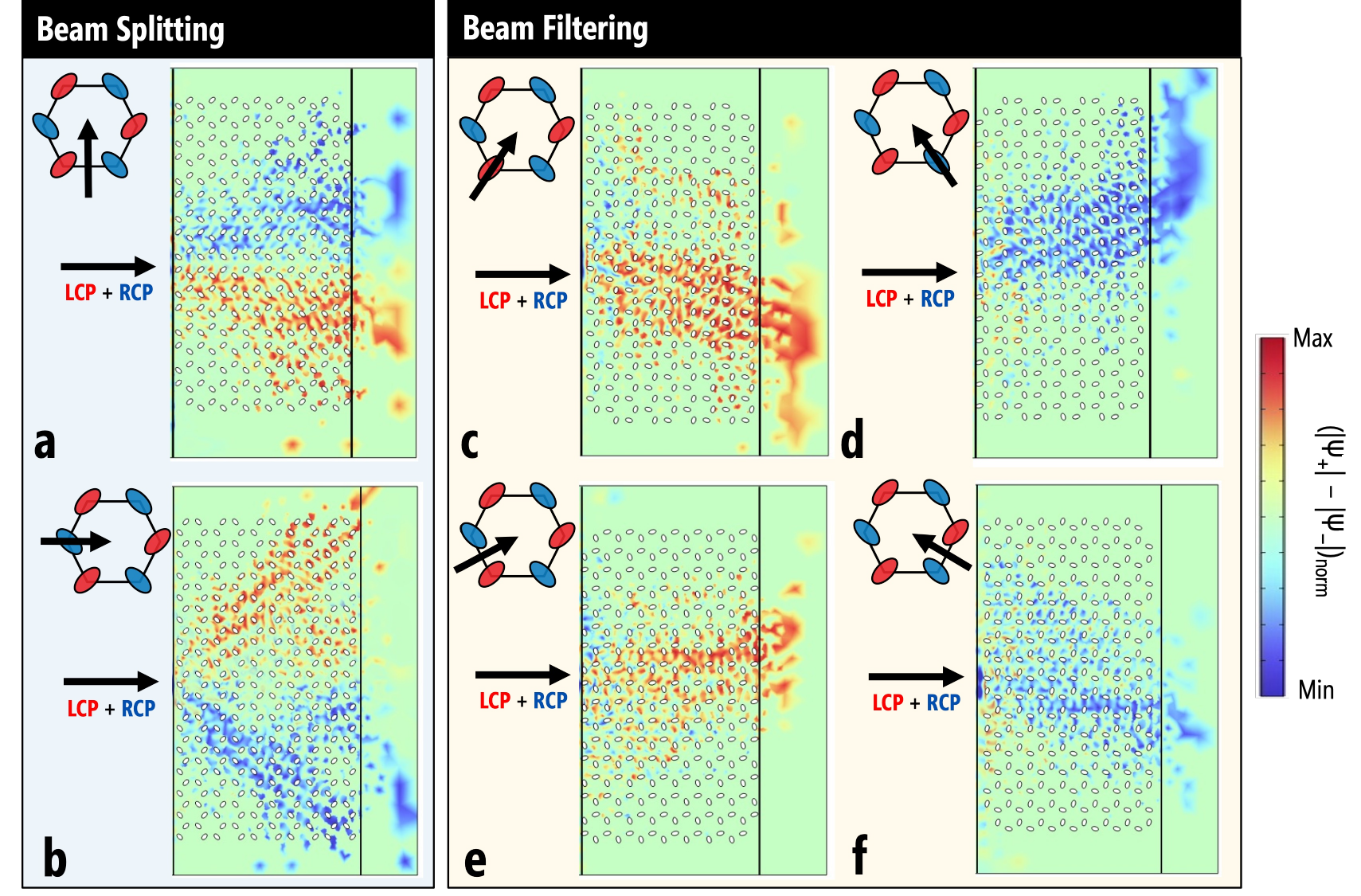}
\caption{
Incidence angle dependence of helicity dependent beam propagation.
The linearly polarized incident beams have a normalized frequency of $f=0.794$, which is different from the normalized frequency used in Fig. 4, $f=0.676$.
(a) and (b) show the beam splitting effect when linearly polarized light is incident from the bottom side and the left side of the hexagonal lattice, respectively.
Compared with (a), the case in (b) exhibits a substantially wider angular range of helicity dependent beam splitting.
(c) and (d) show the helicity dependent filtering effect under oblique upward incidence from the lower left corner and the lower right corner of the structure, respectively.
(e) and (f) are the same as (c) and (d), but with different oblique incident angles, as indicated by the black arrows in the insets.
The markedly different beam propagation profiles show that the helicity dependent beam splitting and filtering effects can be sensitively modulated in the same hexagonal chiral photonic crystal by changing the incident directions.
}
\label{fig:P5}
\end{figure*}

\begin{figure*}[t]
\centering
\includegraphics[width=\textwidth]{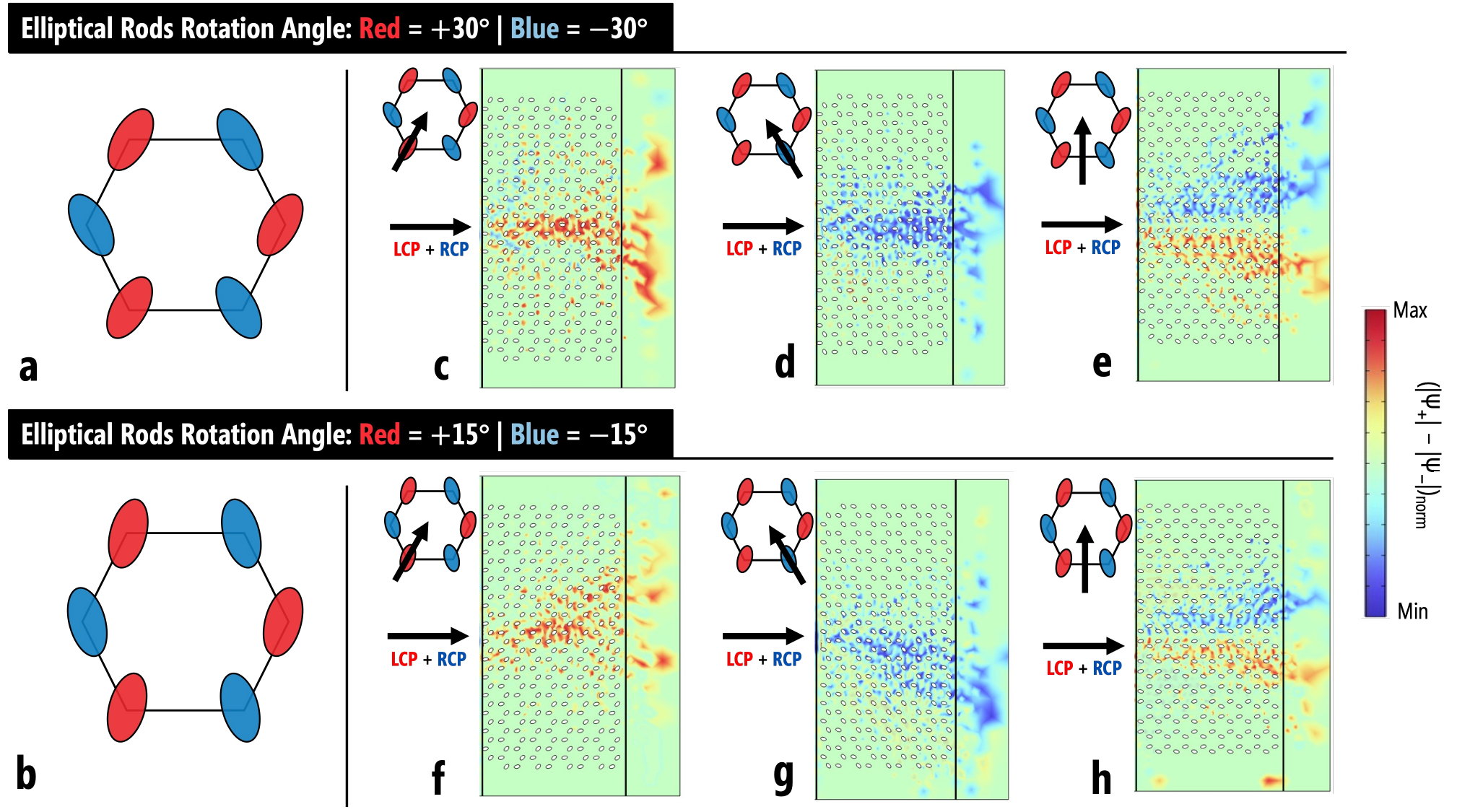}
\caption{
Control of helicity-selective beam propagation through the ellipse rotation angle.
Two representative rotation-angle configurations are considered: (a) $+30^{\circ}$ and $-30^{\circ}$ for the red and blue ellipses, respectively; and (b) $+15^{\circ}$ and $-15^{\circ}$ for the red and blue ellipses, respectively.
Beam-propagation profiles for different incident directions are shown in (c)--(e) for the configuration in (a) and in (f)--(h) for the configuration in (b), with the incident directions indicated by the arrows in the insets.
All incident beams are linearly polarized, and the normalized frequency is $f=0.794$.
(c,d) Helicity-dependent filtering when linearly polarized light is obliquely incident upward from the lower-left and lower-right corners of the structure, respectively, for the $\pm30^{\circ}$ configuration.
(e) Helicity-dependent beam splitting when linearly polarized light is incident from the bottom of the hexagonal lattice for the same configuration.
(f)--(h) Corresponding results for the $\pm15^{\circ}$ configuration shown in (b).
Comparing (c) with (f), and (d) with (g), shows that the propagation directions of the two filtering channels change from oblique propagation to nearly horizontal propagation.
This indicates that the ellipse rotation angle can markedly tune the output direction of the helicity components.
The two opposite helicity components remain separable, but their output directions become closer to each other for the smaller rotation angle, showing that the spatial separation of the two helicity components can also be tuned geometrically.
The ellipse rotation angle therefore provides an effective geometric control parameter for helicity-dependent filtering and beam splitting in the hexagonal chiral photonic crystal.
}
\label{fig:P6}
\end{figure*}

\subsection{Helicity-contrasting light manipulation}

We next examine the conversion of the \(k\)-space helicity splitting into real-space light manipulation in a finite photonic crystal structure. A linearly polarized Gaussian beam is used as the incident source~\cite{Kim2025}. At the source window, the incident field is imposed in the \(E_z\) component with the profile
\begin{equation}
\label{eq:gaussian_incident_field}
E_z^{\mathrm{inc}}(\mathbf{r})
=
E_0
\exp\left(-\frac{u^2}{w_0^2}\right)
\exp\left(i k_b s\right)
\end{equation}
where \(s\) and \(u\) are the coordinates parallel and perpendicular to the incident direction, respectively. Here, \(E_0\) is the incident amplitude, \(w_0\) is the Gaussian beam waist, and \(k_b\) is the wave number in the background medium. The background region, with $\varepsilon_0 = \mu_0 = 1$, gives \(k_b = 2\pi f/a_0\) for the normalized frequency \(f = \omega a_0/(2\pi c)\). In the numerical model, the Gaussian profile is applied within a finite source window, whose width sets the effective aperture, while \(w_0\) controls the transverse beam width.

After entering the altermagnetic photonic crystal, the two helicity components of the linearly polarized beam couple to Bloch modes of different helicity, set by momentum matching at the interface: the tangential component of the incident wave vector selects which modes are excited, and the energy then flows along the group velocity, which is normal to the IFC at the coupling point. When the two helicities couple to modes with different group-velocity directions, the linearly polarized input splits into two spatially separated helicity channels. 

Figure 4(a) shows four distinctive cases of beam propagation, labeled I to IV, when linearly polarized light impinges on the altermagnetic photonic crystal at different incident directions at a normalized frequency $f=0.676$. In the helicity-resolved IFC [Fig. 4(a)], the red and blue contours denote the LCP and RCP branches, respectively. The black arrows represent incident wave vector directions, while the red and blue arrows represent the group velocity directions of the corresponding momentum-matched LCP and RCP propagating modes in the photonic crystal, respectively. The propagation maps of cases I to IV demonstrate four distinct beam splitting and filtering functions enabled by the finite altermagnetic photonic crystal slab, namely: (i) Case I: Beam splitting of LCP and RCP; (ii) Case II: Same as (i) but with opposite spatial splitting; (iii) Case III: LCP is transmitted while RCP is blocked; and (iv) Case IV: RCP is transmitted while LCP is blocked. The \emph{beam splitting} effect in Cases I and II arises from the directionally-contrasting group velocities of the LCP and RCP branches. The \emph{beam filtering} effect in Cases III and IV arises due to the presence of only a single momentum-matched helicity channel. 

Figure 4(b) summarizes these regimes: the gray sector marks the incident angles where both helicity branches are accessible and beam splitting occurs, while the red and blue sectors mark the angles where only the LCP or RCP branch is transmitted. The same mirror-symmetric chiral texture is thus switched between a beam splitter and a helicity filter purely by varying the incident direction, without changing any structural parameter.

We further examine the beam splitting and filtering effects at a different isofrequency of $f=0.794$ in Fig. 5. 
Figures 5(a) and 5(b) how beam splitting under two incident directions, indicated by the bold arrows in the insets. The incident direction strongly shapes the splitting: relative to Fig.~5(a), the direction in Fig.~5(b) yields a larger spatial separation of the LCP and RCP beams and reverses their separation directions. Similarly, the beam filtering characteristics are also sensitively influenced by the \emph{oblique} incidence directions [Figs. 5(c) to 5(f)]. In Figs.~5(c) and 5(e) the output is LCP-dominated, whereas mirror-reversing the incident direction [Figs.~5(d) and 5(f)] switches the output to RCP.

\subsection{Quantification of output-helicity selectivity and transmission efficiency}

We employ the RS field intensities of the two helicity components, $\psi_\pm$, to quantify the beam filtering efficiency of the finite altermagnetic photonic crystal slab. The $\psi_\pm$ are integrated separately over the left incident region and the right output region. The right output region and the left incident region are denoted by $\Omega_{\mathrm{out}}$ and $\Omega_{\mathrm{in}}$, respectively. For LCP and RCP filtering, the selected output components are \(\psi_{+}\) and \(\psi_{-}\), respectively. The output-helicity fraction can be defined as:
\begin{equation}
\label{eq:S_pm}
\Xi_{\pm}
=
\frac{
\int_{\Omega_{\mathrm{out}}} |\psi_{\pm}|^{2} \, \mathrm{d}A
}{
\int_{\Omega_{\mathrm{out}}}
\left(
|\psi_{+}|^{2}
+
|\psi_{-}|^{2}
\right)
\, \mathrm{d}A
}
\end{equation}
where $\Omega_\mathrm{out}$ represents the output region. Here, $\Xi_{\pm} \to 1$ indicates stronger helicity filtering effect. In addition to the filtering capability, the amount of light that emerges from the output region is another important metric for assessing the performance of the altermagnetic photonic crystal slabs in helicity beam filtering. We define the normalized output intensity $T_{\mathrm{out}}$ as the ratio of the total RS field intensity integrated over the right output region to that integrated over the left incident region:
\begin{equation}
\label{eq:T_out}
T_{\mathrm{out}} =
\frac{
\int_{\Omega_{\mathrm{out}}} \left( |\psi_{+}|^{2} + |\psi_{-}|^{2} \right) \, \mathrm{d}A
}{
\int_{\Omega_{\mathrm{in}}} \left( |\psi_{+}|^{2} + |\psi_{-}|^{2} \right) \, \mathrm{d}A
}
\end{equation}
where $T_{\mathrm{out}}$ is obtained from the integrated RS field intensities and is used to compare the relative output intensities of the finite altermagnetic photonic crystal slab. 

Using Eqs. (7) and (8), the LCP helicity filtering in Fig.~5(c) yields $\Xi_{+}=0.857$ and $T_{\mathrm{out}}=0.803$, while the RCP helicity filtering in Fig.~5(d) yields $\Xi_{-}=0.888$ and $T_{\mathrm{out}}=0.817$. This balance is essential: a high target-helicity fraction is meaningful only when the selected channel also retains sufficient output intensity, as achieved simultaneously in Figs.~5(c) and 5(d).

The target-helicity fractions $\Xi_\pm > 0.85$ demonstrate that the
proposed altermagnetic photonic crystal operates as an effective
helicity filter, while the normalized output intensities
$T_{\mathrm{out}} \sim 0.8$ confirm that the selected channel retains
appreciable output strength. Notably, the incident beams in
Figs.~5(c) and 5(d) are linearly polarized, each an equal
superposition of LCP and RCP components. The high helicity purity of the
output therefore originates from the momentum-matched coupling and
helicity-resolved propagation of the structure itself, rather than from
any specific LCP/RCP ratio in the incident beam. This relaxes the
requirement for external polarization preparation and makes the incident
direction itself a practically useful channel-selection parameter for engineering the beam splitting and filtering performance of the altermagnetic photonic crystal.

\subsection{Beam splitting and filtering characteristics tuning via elliptical element geometry}
 
We now show that the beam splitting and filtering effects can also be tuned by varying the rotation angles of the elliptical elements. In Fig. 6, the alternating distribution of positive and negative chiral ellipses and the mirror-symmetric arrangement at the hexagonal vertices are kept unchanged. Only the orientation of the elliptical major axes is modified. 
 
Two representative configurations are considered, with rotation angles of $\pm30^{\circ}$ and $\pm15^{\circ}$ for the red and blue ellipses, respectively [Figs. 6(a) and 6(b)]. For the $\pm30^{\circ}$ configuration, Figs. 6(c) and 6(d) show that complementary oblique incident directions can still select LCP and RCP outputs, respectively. This indicates that the helicity filtering function is preserved after changing the ellipse angle. Compared with the $\pm45^{\circ}$ case, the light beam propagates almost normal to the incident interface in the photonic crystal in the $\pm30^{\circ}$ cases because the IFC normal directions have been reoriented by the geometric rotation. Figure 6(e) shows the beam splitting with bottom incidence for the same $\pm30^{\circ}$ configuration. The LCP and RCP branches similarly exhibit spatial separation, but with a different separation profile as compared with the $\pm45^{\circ}$ cases. 
 
We further reduce the rotation angle to $\pm15^{\circ}$. The spatial profile of the beam splitting [Figs. 6(f) and 6(g)] is again modulated as compared with the $\pm30^{\circ}$ and $\pm45^{\circ}$ cases. In Fig. 6(h), the LCP and RCP beam splitting via bottom incidence remains identifiable, but the two helical beams at the output region become closer to each other, suggesting that an overly small ellipse rotation angle weakens the beam splitting effect. The helicity-contrasting beam modulation is thus not caused by the presence of `accidental' propagation channels at a specific geometry, but more generally arises from the helicity-resolved dispersion, where geometrical rotation modifies the IFC contours and hence the output beam profiles.

Figures 4 to 6 jointly reveal that helicity beam manipulation is governed by several coupled factors: (i) The operating frequency selects the relevant isofrequency contour, or equivalently the set of Bloch momentum channels available for propagation; (ii) The ellipse orientation modifies the IFC shape and helicity distribution, thereby controlling the group-velocity directions and the separation angle between the two helicity channels; and (iii) The incident direction determines which portion of the helicity-resolved IFC is accessed through momentum matching. The interplay among these factors thus provides multiple tuning knobs that can be jointly optimized for efficient helicity-dependent beam manipulation in a finite-structure altermagnetic photonic crystal slab.

\section{Conclusion}

In summary, we proposed a mirror-symmetry-enforced photonic altermagnet based on a hexagonal chiral photonic crystal. By arranging positive-and negative-chirality elliptical elements at the vertices of a regular hexagon, the structure maintains zero average chirality while breaking chirality-reversal, $\mathcal{P}\mathcal{T}$-type, and translation restoration. The remaining mirror relation constrains the two helicity branches and gives rise to mirror-related helicity splitting in momentum space. Full-wave simulations show that the helicity splitting is controlled by the coupling between material chirality and ellipse orientation. The $\pm45^\circ$ configuration produces clear helicity-resolved band and isofrequency-contour splitting, while reversing the ellipse orientation exchanges the two helicity channels. Riemann--Silberstein field analysis further confirms that the helicity-contrasting branches originate from the optical helicity of the eigenmodes rather than from material-domain labeling.

Using a finite-structure altermagnetic photonic crystal, we show that a linearly polarized Gaussian beam can be separated into two helicity-dependent paths or selectively filtered into a dominant helicity output. The filtering cases achieve target helicity fractions above 0.85 with normalized output intensities close to 0.8. In addition, changing the incidence interface angle and the elliptical element rotation angle can tune the output direction and channel separation, thus providing a geometric control knob for helicity routing. Our findings extend photonic altermagnetism to a mirror-symmetric hexagonal lattice, broadening it beyond the square-lattice, rotation-enforced paradigm. The proposed altermagnetic photonic crystal offers a practical route toward on-chip helicity-resolved beam splitting, filtering, and routing \cite{turner2013miniature, voronin2022single}.

\begin{acknowledgements}
This work is supported by the Scientific Research Fund of Hunan Provincial Education Department with Grants No. 25C0072, and the Provincial Natural Science Foundation of Hunan with Grants No. 2026JJ50371. Y.S.A acknowledges the supports from the Kwam Im Thong Hood Cho Temple Early Career Chair Professorship and the Singapore Ministry of Education (MOE) Academic Research Fund (AcRF) Tier 2 Grant (MOE-T2EP50224-0021).

\end{acknowledgements}

\section*{Author Declarations}

\subsection*{Conflict of Interest}
\noindent The authors declare that there are no conflicts of interest.

\section*{Data Availability}
The data that support the findings of this study are available from the corresponding author upon reasonable request.

\bibliography{ref}

\end{document}